\documentclass{article}
\usepackage{etex}
\usepackage[T1]{fontenc}
\usepackage[latin1]{inputenc}
\usepackage{graphicx}
\usepackage{amsfonts}
\usepackage{units}
\usepackage{float}
\usepackage{easybib}
\usepackage[intlimits]{amsmath}
\usepackage{tabularx}

\newcommand{\beq}{\begin{eqnarray}}
\newcommand{\eeq}{\end{eqnarray}}
\newcommand{\bdm}{\begin{displaymath}}
\newcommand{\edm}{\end{displaymath}}
\newcommand{\be}{\begin{equation}}
\newcommand{\ee}{\end{equation}}

\newcommand{\pd}[1]{\, \partial #1 \,}
\newcommand{\td}[1]{\, \mathrm{d} #1 \,}
\newcommand{\intl}{\int\limits}

\newcommand{\HF}[1]{\; H\left[ #1 \right]}  
\newcommand{\DF}[1]{\; \delta\left( #1 \right)}  
\newcommand{\nup}[1]{\nu^{\prime #1}}
\newcommand{\SED}{f^{\prime}}
\newcommand{\ergcm}{\;\hbox{erg}\:\hbox{cm}^{-3}}

\newcommand{\epec}{\epsilon_{ec}}

\title{Modelling of blazar SEDs with the nonlinear SSC cooling process}
\author{M. Zacharias \& R. Schlickeiser}
\date{}
\begin{document}
\maketitle
\begin{abstract}
Observations of blazar flaring states reveal remarkably different variability time scales. Especially rapid flares with flux doubling time scales of the order of minutes have been puzzling for quite some time. Many modeling attempts use the well known linear relations for the cooling and emission processes in the jet in a steady-state scenario, albeit the obvious strongly time-dependent nature of flares. Due to the feedback of self-produced radiation with additional scattering by relativistic electrons, the synchrotron-self Compton (SSC) effect is inherently time-dependent. Although this feedback is usually implemented in numerical treatments, only recently an analytical analysis of the effects of this nonlinear behaviour has been performed. Here, we report our results concerning the effect of the time-dependent SSC on the spectral energy distribution (SED) of blazars. We calculated analytically the synchrotron and the SSC component, giving remarkably different spectral features compared to the standard linear approach. Adding an external photon field to the original setting, we could implement quite easily the effect of an additional external Compton (EC) cooling, since such strong external photon fields are observed in flat spectrum radio quasars (FSRQ), a subclass of blazars. Calculating the resulting flux due to the EC cooling, we were able to show that the resulting inverse Compton component strongly depends on the free parameters, and that SSC could potentially have a strong effect in FSRQs, contrary to what is usually assumed.
\end{abstract}

%
%

\section{Introduction}

Blazars, as a class of Active Galactic Nuclei (Urry \& Padovanni 1995), are characterized by strong variability in all wave-bands. Additionally, a broad range of variability time scales is observed, ranging from minutes, as in the cases of PKS 2155-304 (Aharonian et al. 2007) and PKS 1222+216 (Tavecchio et al. 2011), to months (e.g. Abdo et al. 2010). Especially the very short variability time scales are puzzling, since they imply emission regions to be much smaller than the event horizon of the black hole, which should be a lower limit on the jet width.

Under such circumstances it seems quite surprising that most modeling attempts still use the usual linear approach in a steady-state scenario (Ghisellini et al. 2009). Especially rapid flares are far from steady-state, and need a real time-dependent treatment. Numerical studies normally incorporate time-dependent features by default. There is one process, where this is particularly important: the synchrotron-self Compton effect (SSC, Jones, O'Dell \& Stein 1974). Under the influence of a magnetic field relativistic electrons generate synchrotron radiation, which is afterwards up-scattered to very high energies by the same electrons. Since the electrons naturally lose energy during this process the energy density in the synchrotron field will decrease over time, and so will the efficiency of the SSC process. Since numerical treatments calculate the synchrotron and SSC radiation after every time step, this time-dependency is implemented almost unintentionally. 

Only recently Schlickeiser (2009) calculated analytically the fully time-dependent cooling term for the SSC radiation, which depends on an integral over the electron distribution function itself. The differential equation (Kardashev 1962)
\be
\frac{\pd{n(\gamma,t)}}{\pd{t}} - \frac{\pd}{\pd{\gamma}}\left[ |\dot{\gamma}_{tot}|n(\gamma,t) \right] = S(\gamma,t) ,
\label{PDEn1}
\ee
describing the behaviour of the electron distribution $n(\gamma,t)$ under the influence of cooling described by $\dot{\gamma}_{tot}$ and the source $S(\gamma,t)$, then becomes nonlinear.

Due to the time dependency the rate of the SSC cooling decreases over time. This implies that, although the SSC cooling might be stronger in the beginning than the linear synchrotron cooling, the synchrotron cooling eventually becomes stronger than the SSC cooling. This can be incorporated by using both cooling terms in the differential equation (\ref{PDEn1}), giving
\beq
|\dot{\gamma}_{tot}| = |\dot{\gamma}_{syn}|+|\dot{\gamma}_{ssc}| = D_0\gamma^2 + A_0\gamma^2 \intl_0^{\infty}\td{\gamma} \gamma^2 n(\gamma,t) ,
\label{totcoolrate}
\eeq
where $\gamma$ is the Lorentz factor of the electrons, $D_0 = 1.256\times 10^{-9} b^2$s$^{-1}$, $A_0 = 1.15\times 10^{-18} R_{15}b^2$cm$^{3}$s$^{-1}$, with a magnetic field of strength $B=b$Gauss, and a radius of the emission blob $R=10^{15} R_{15}$cm.

In section 2 we will present the solution to the differential equation (\ref{PDEn1}), which depends on the injection parameter $\alpha$, which is the ratio of SSC to synchrotron cooling at time of injection. We will also briefly discuss the implications of this parameter in section 2. Section 3 will present some example spectral energy distributions (SED) showing the significant differences between the linear and the nonlinear approach. In section 4 we will give examples of the inclusion of external photon fields into the cooling scenario, again altering the results. We will conclude in section 5.

%
%

\section{The injection parameter}

Schlickeiser, B\"ottcher \& Menzler (2010) used the cooling term (\ref{totcoolrate}) and a source term $S(\gamma,t) = q_0\DF{\gamma-\gamma_0}\DF{t}$ to solve the differential equation (\ref{PDEn1}). As stated above, the solution strongly depends on the injection parameter $\alpha$, giving for $\alpha\ll 1$
\begin{equation}
n(\gamma ,t,\alpha \ll 1)= q_0 \HF{\gamma _0-\gamma} \DF{\gamma -\frac{\gamma _0}{1+D_0\gamma _0t}},
\label{b2}
\end{equation} 
which is solely determined by the linear synchrotron cooling. 

For $\alpha\gg 1$ the solution is divided into two parts for early and late times. For early times one finds
\begin{equation}
n(\gamma ,\gamma _0,t\le t_c,\alpha \gg 1) = q_0 \HF{\gamma _0-\gamma} \HF{t_c-t} \DF{\gamma -\frac{\gamma _0}{(1+3A_0q_0\gamma _0^3t)^{1/3}}} ,
\label{b3}
\end{equation}
reflecting the nonlinear cooling. For later times
\begin{equation}
n(\gamma ,\gamma _0,t\ge t_c,\alpha \gg 1) = q_0 \HF{\gamma _B-\gamma} \HF{t-t_c} \DF{\gamma -\frac{\gamma _B}{\frac{1+2\alpha ^3}{3\alpha ^3}+D_0\gamma_Bt}} ,
\label{b4} 
\end{equation}
which is a modified linear solution. The critical time $t_c=(3\gamma _BD_0)^{-1}$, and $\gamma_B = \gamma_0/\alpha$.

We call $\alpha$ the injection parameter, since it is defined as the square-root of the ratio of the non-linear cooling term to the linear cooling term at time of injection, i.e.
\beq
\alpha^2 = \frac{|\dot{\gamma}_{ssc}(t=0)|}{|\dot{\gamma}_{syn}|} = \frac{A_0q_0\gamma_0^2}{D_0} .
\label{alpha1}
\eeq
For $\alpha\ll 1$ we see that the linear cooling dominates, resulting in the solution (\ref{b2}). If $\alpha\gg 1$, the non-linear cooling at least initially dominates, giving the solutions (\ref{b3}) and (\ref{b4}) for early and late times, respectively. The time $t_c$ marks the transition from non-linear to linear cooling.

Since the Compton dominance, i.e. the ratio of the peak values of the two components in the SED, should be related to the ratio of the cooling terms, we expect a dominating synchrotron component for $\alpha\ll 1$, while for $\alpha\gg 1$ the SSC component should show higher fluxes.

%
%

\section{Example SEDs}

\begin{figure}
\begin{minipage}[t]{0.49\linewidth}
\centering \resizebox{\hsize}{!}
{\includegraphics{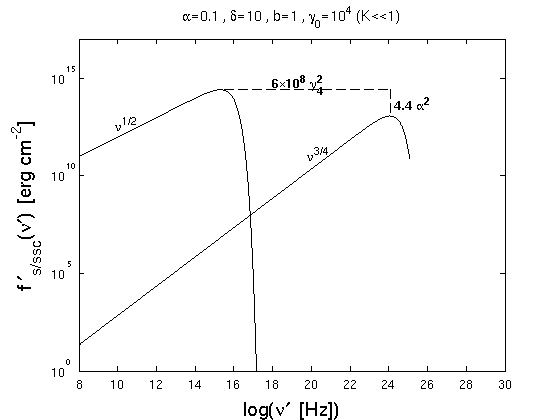}}
\end{minipage}
\hspace{\fill}
\begin{minipage}[t]{0.49\linewidth}
\centering \resizebox{\hsize}{!}
{\includegraphics{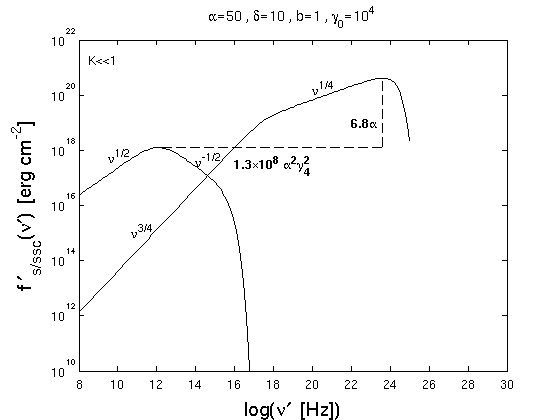}}
\end{minipage} 
\caption{{\it Left:} SED for $\alpha\ll 1$ in the Thomson-limit. {\it Right:} SED for $\alpha\gg 1$ in the Thomson-limit. Parameters for each case are given at the top. Indicated are the Compton dominance (vertical line) and ratio of the peak frequencies (horizontal line), as well as the power-laws.}
\label{fig01}
\end{figure} 

In figure \ref{fig01} we present two example SEDs $\SED(\nup{})$. They were calculated by Zacharias \& Schlickeiser (2012a) by first calculating the intensity spectrum for each case and then integrating these spectra with respect to time, giving the fluence spectra. The SED follows by multiplying the fluence with the frequency $\nu$. With the help of the Doppler factor $\delta$ we transformed our spectra to the system of rest of the host galaxy (primed quantities). 

The left panel in figure \ref{fig01} shows the case for $\alpha\ll 1$, and one can clearly see that the synchrotron component dominates the SSC component. 

The right panel shows the case for $\alpha\gg 1$, and here the SSC component dominates the synchrotron component as expected. Secondly, both components exhibit broken power-laws, which are a natural result of the nonlinear cooling. We note the important fact that these breaks are due to the cooling term and do not depend on the underlying electron distribution (c.f. Zacharias \& Schlickeiser 2010 for the synchrotron component). Thus, we can explain the breaks in the SEDs without the need for fancy electron distributions, which are often given without any theoretical justification (e.g. Abdo et al. 2011).

%
%

\section{External photon fields}

Especially flat spectrum radio quasars (FSRQs) show strong thermal components, albeit the beamed nonthermal radiation. Thus, the electrons of these sources could also be cooled by inverse Compton cooling with these thermal components, like photons directly from the accretion disk (Dermer \& Schlickeiser 1993), from the broad line regions (Sikora et al. 1994) or the dusty torus (Blazejowski et al. 2000, Arbeiter et al. 2002). 

Zacharias \& Schlickeiser (2012b) included this type of inverse Compton cooling into the above given model by adding the inverse Compton cooling term
\beq
|\dot{\gamma}_{ec}| \approx 4.4\times 10^{-8} \left( \frac{u^{\prime}_{ec}}{\ergcm} \right)\Gamma_b^2\gamma^2 \,\hbox{s}^{-1} , 
\label{exlossrate}
\eeq
to the cooling rate (\ref{totcoolrate}). Here $u^{\prime}_{ec}$ is the energy density of the external photon field in the galactic frame, and $\Gamma_b$ is the Lorentz factor of the plasma blob in the jet.

\begin{figure}
\begin{minipage}[t]{0.49\linewidth}
\centering \resizebox{\hsize}{!}
{\includegraphics{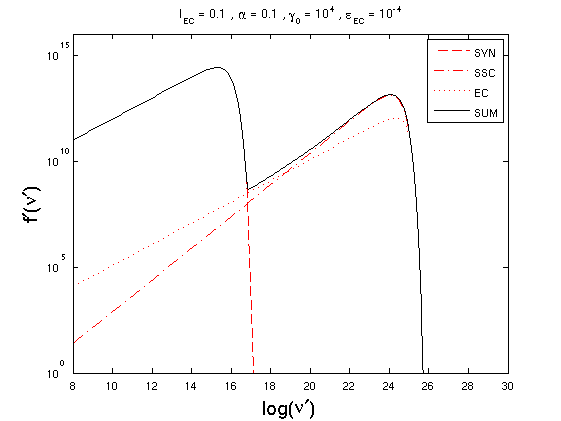}}
\end{minipage}
\hspace{\fill}
\begin{minipage}[t]{0.49\linewidth}
\centering \resizebox{\hsize}{!}
{\includegraphics{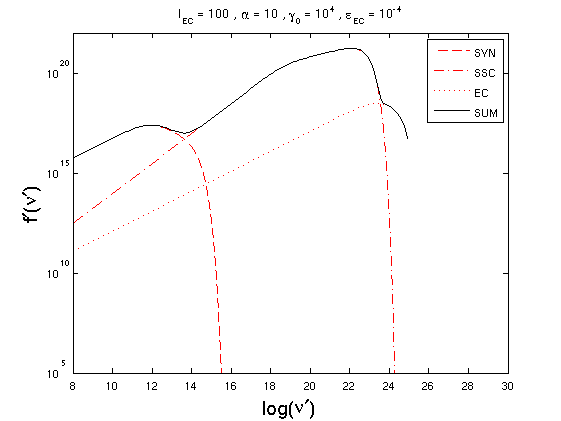}}
\end{minipage} 
\caption{{\it Left:} SED for $\alpha\ll 1$ and $l_{ec}\ll 1$. {\it Right:} SED for $\alpha\gg 1$ and $l_{ec}\gg 1$. Parameters for each case are given at the top.}
\label{fig02}
\end{figure}

After calculating the resulting external Compton intensity and fluence, Zacharias \& Schlickeiser (2012b) were able to present the complete SEDs, of which two examples are given in figure \ref{fig02}.

The plots depend now also on the relative strength of the linear cooling terms $l_{ec} = |\dot{\gamma}_{ec}| / |\dot{\gamma}_{syn}|$, and the normalized external photon energy $\epec$ (assuming a monochromatic photon source).

The main result is that the final SED depends strongly on the free parameters, and that both a dominating SSC or EC component are possible. Zacharias \& Schlickeiser (2012b) argue, therefore, that the SSC component should not be neglected in blazar SEDs, and the correct time-dependent treatment should be applied.

%
%

\section{Conclusions}

In this proceeding we have given the results of recent analytical investigations concerning the time-dependent nature of the SSC process and the consequences for blazar SED modeling attempts. We first presented the results of Zacharias \& Schlickeiser (2012a), in which we calculated the SED including the synchrotron and the SSC component. Then we proceeded with the recent work of Zacharias \& Schlickeiser (2012b), where we added the effect of external Compton cooling, yielding a third broad component in the SED.

To conclude, we can say that our investigation has clearly shown the importance of the inclusion of the time-dependent nature of the SSC cooling term, especially in blazar flaring states, which are far from steady-state. The resulting spectra differ significantly from the usual linear approach, where these effects are not taken into account. Secondly, due to the broad range of free parameters in the EC model, the SSC component should not be dismissed beforehand while modeling blazars. As we were able to show here, the SSC component might be as important as the EC component, if the correct time-dependent treatment is applied, which is especially necessary in flaring states.

We aim to calculate the light-curves and the optical depths in a future work in order to further highlight the importance of the time-dependent SSC cooling treatment.

%
%

\end{document}